\documentclass[review, 1p]{elsarticle}
\usepackage{graphicx,amsmath,amssymb,color,multirow,doi}

\usepackage{courier}

\definecolor{forest}{rgb}{0.04, 0.39, 0.13}

\definecolor{bittersweet}{rgb}{1.0, 0.44, 0.37}

\newcommand\etal{\textit{et al. }}

\begin{document}
\begin{frontmatter}
\title{Network analysis reveals news press landscape and asymmetric user polarization}
\author{Byunghwee Lee} 
\address{Center for Complex Networks and Systems Research,
Luddy School of Informatics, Computing, and Engineering, Indiana University, Bloomington, Indiana 47408, USA}
\address{Institute of Basic Science, Sungkyunkwan University, Suwon 16419, Republic of Korea}
\author{Hyo-sun  Ryu}
\address{Institute of Technology \& Democracy, Sungkyunkwan University, Seoul 03063, Republic of Korea}
\author{Jae Kook Lee}
\address{Department of Media and Communication, Sungkyunkwan University, Seoul 03063, Republic of Korea}
\author{Hawoong Jeong}
\address{Department of Physics, Korea Advanced Institute of Science and Technology, Daejeon 34141, Republic of Korea}
\author{Beom Jun Kim}
\ead{Corresponding author: beomjun@skku.edu}
\address{Department of Physics, Sungkyunkwan University, Suwon 16419, Republic of Korea}

\date{\today}

\begin{abstract}
Unlike traditional media, online news platforms allow users to consume content that suits their tastes and to facilitate interactions with other people. However, as more personalized consumption of information and interaction with like-minded users increase, ideological bias can inadvertently increase and contribute to the formation of {\em echo chambers}, reinforcing the polarization of opinions.  Although the structural characteristics of polarization among different ideological groups in online spaces have been extensively studied, research into how these groups emotionally interact with each other has not been as thoroughly explored. From this perspective, we investigate both structural and affective polarization between news media user groups on Naver News, South Korea's largest online news portal, during the period of 2022 Korean presidential election. By utilizing the dataset comprising 333,014 articles and over 36 million user comments, we uncover two distinct groups of users characterized by opposing political leanings and reveal significant bias and polarization among them. Additionally, we reveal the existence of echo chambers within co-commenting networks and investigate the asymmetric affective interaction patterns between the two polarized groups. Classification task of news media articles based on the distinct comment response patterns support the notion that different political groups may employ distinct communication strategies. Our approach based on network analysis on large-scale comment dataset offers novel insights into characteristics of user polarization in the online news platforms and the nuanced interaction nature between user groups. 
\end{abstract}

\begin{keyword}
\texttt{Polarization, Echo chamber, Affective polarization, News media, Social media}
\end{keyword}

\end{frontmatter}

\section{Introduction}
In the information age, online news platforms and social media have become more pivotal in shaping public discourse than at any other time introducing novel avenues for information dissemination and interaction. The transition from traditional media to digital platforms has not only altered the way information is disseminated but also how it is consumed and interacted with~\cite{robertson2023users, cinelli2021echo, bode2016political, del2016spreading, gonzalez2023asymmetric, guess2023social, guess2023reshares, kwak2021frameaxis}. For instance, online news platforms are offering users the immediate and cost-free opportunity to access information from a variety of news media according to their preferences. Also, these news platforms have enabled individuals to share of varied opinions on specific issues with others through various user participation channels such as comments, votes, and re-sharing of contents. 

However, there is ongoing debate about whether the online social platforms aid in harmonizing diverse voices or, conversely, amplify cohesion among like-minded individuals, potentially fostering increased bias against groups with differing opinions and leading to the formation of {\em echo chambers}~\cite{cinelli2021echo, cota2019quantifying}  or {\em filter bubbles}~\cite{robertson2023users, flaxman2016filter, bakshy2015exposure}. This issue becomes particularly pronounced during significant political events, where opinion polarization can have far-reaching consequences.

Researchers have probed users' political inclinations and polarization on social media~\cite{hohmann2023quantifying, adamic2005political}, with studies like Cota \etal~\cite{cota2019quantifying} using Twitter network to gauge echo chamber effects in information spreading over political communication networks. Similarly, Cinelli \etal~\cite{cinelli2021echo} analyzed collective political alignment of the media sources associated with users’ shared content, providing insights into the structure of political discourse network among users. Gonzalez \etal analyzed Facebook data during the 2020 US election and revealed ideological segregation and an asymmetry in audience engagement, emphasizing dominance of conservative information in the news ecosystem~\cite{gonzalez2023asymmetric}.  
Recent studies have further illuminated the intricate link between news sharing behavior on online platforms and users' political leanings~\cite{del2025evaluating, del2024analyzing, cicchini2022news}. Investigations into user behavior during political events have shown that individuals tend to predominantly share news content that aligns with their pre-existing political beliefs, even when exposed to diverse sources.

In addition, researchers proposed models of opinion dynamics on social network to understand the mechanisms by which extreme opinions are formed from users' moderate initial conditions~\cite{baumann2020modeling, mas2013differentiation}. Recent literature has incorporated cognitive aspects of human beliefs into models integrated with the influence of social network to model individual's belief change that leads to group level polarization or consensus of a social group~\cite{galesic2021integrating, dalege2022using}.

Although prior research has extensively studied polarization on social media, most studies have emphasized structural aspects, such as network topology and patterns of content sharing, with relatively less attention given to the affective or emotional dimensions of user interactions~\cite{hohmann2023quantifying, kubin2021role, iyengar2019origins}. Recent studies have begun to explore affective polarization through explicit user reactions, such as likes and content resharing behaviors associated with polarized information in social media~\cite{cinelli2021echo, gonzalez2023asymmetric}. However, a critical gap remains regarding the direct emotional exchanges between users embedded within structured comment-response interactions, particularly in the context of online news media platforms. Moreover, many existing analyses rely heavily on externally defined political categorizations of media outlets, which may not be available or reliable in all contexts. Thus, it remains unclear whether distinct ideological groups and their patterns of affective polarization can be identified solely based on intrinsic user interaction data, without predefined ideological labels.


In this study, we address these gaps by employing concepts and methodologies of network science to quantitatively analyze reaction-based affective interaction patterns, particularly focusing on explicit sympathy (likes) and antipathy (dislikes), between ideologically opposed user groups on an online news platform. Specifically, we investigate whether polarized user groups can be identified purely from collective reaction patterns to user comments on articles from various news media outlets, independent of externally defined political leanings. By analyzing both structural and emotional dimensions of interactions using a comprehensive dataset from a major South Korean news portal, we aim to uncover distinctive patterns of affective polarization and communication strategies among user groups, providing novel insights into the emotional dynamics underlying news media polarization.


Particularly, we concentrate on `Naver News', South Korea's largest online news portal, during a critical period of the 2022 presidential election of South Korea. This period offers a unique opportunity to examine user behaviors in relation to articles with different perspective over a variety of news media. By analyzing a comprehensive dataset of over 333,014 articles and 36 million user comments, we endeavor to uncover patterns of bias and polarization in both media and user engagement, and affective interactions between divergent user groups.

We begin by examining how different news media are clustered based on user responses, highlighting distinct media groups with contrasting ideologies. Next, we explore the distribution of political leanings of users and how it is correlated with user activities. Through an analysis of the link structure within the co-commenting network of users, we uncover the presence of {\em echo chambers} within the online news platform. Additionally, we examine asymmetric interaction patterns between polarized groups revealing distinct characteristics of affective polarization in digital discourse. Finally, drawing from our observations of asymmetric affective response patterns between these groups, we conduct a classification task predicting the political leanings of news media articles, thereby confirming the utilization of distinct communication strategies by contrasting user groups. 

This research contributes to the broader understanding of bias in news consumption and the emergence of both ideological and affective polarization within online news platforms. By examining these phenomena in the context of a major political event, we provide valuable insights into the shaping of public opinion and political discourse in the digital age.

\section{Dataset}
In this study, we utilize a large-scale news data from the largest South Korean news portal Naver News during the period of 2022 presidential election of South Korea between the year of 2021 and 2022. To ensure a comprehensive collection of articles from various media on the same topic, articles related to presidential candidates and corresponding comments were collected over a span of approximately six months leading up to the election day (from September 1, 2021, to March 8, 2022, totaling 189 days). Specifically, articles and comments related to the four major presidential candidates ``Jae-myung Lee / Suk-yeol Yoon/ Cheol-soo Ahn / Sang-jung Sim'' were gathered with the query term ``Candidate'' in addition to the candidate names. We collected the URLs of news articles whose titles or contents included these queries. The news data were collected between June 14, 2022 to July 1, 2022 and the comments data were collected between September 14, 2022 to October 4, 2022. Therefore, there is a time gap of approximately six months between the last posting date of comments and the comment collection date. We acknowledge this time gap, during which some emotionally charged or harmful comments may have been deleted, either by user action or through platform moderation algorithms. Nevertheless, our analysis robustly detects clear patterns of both structural and affective polarization, indicating that the key findings remain valid even if comments from more ideologically extreme users were disproportionately removed.

The collected news article data related to the four candidates is composed of 333,014 articles from a total of 102 different news media. In Naver News platform, users can write comments on articles and other users can respond to these comments through `replies' or by voting `sympathy/antipathy' to the comments, similarly to like/dislike in other social media platforms. A total of 36,925,124 comments with the response data including the number of replies and the number of sympathy/antipathy were also collected.  

In summary, our dataset comprises the two main parts: news-level information and comment-level information. The news-level information includes original data about the news content, such as the news head, body, news media (name of publisher), publication timestamp, and comment counts. The comment-level information includes an anonymous unique identifier (ID) to identify users consistently across comments, the comment's timestamp, and information about user reactions to the comment, including reply count and sympathy and antipathy count.

\section{Results}

\subsection{Clustering of news media based on comment responses reveals two contrasting media groups}

Before conducting the analysis, we hypothesize that user engagement (including comments, reply, and expressions of sympathy or antipathy) on articles from a specific media outlet mainly reflect the political orientation of that outlet's readership. Generally, news consumers tend to favor sources that echo their political beliefs, often engaging more with content that aligns with their views and showing greater trust in sources reflecting their political stance. This pattern is well-documented in several studies~\cite{jurkowitz2020us, stroud2018american, frimer2017liberals, iyengar2008selective, iyengar2009red, nam2013not}.

On platforms like Naver News, where readerships of different news media are not distinctly separated, users encounter a feed featuring articles from various sources. They can interact with these articles by commenting or reacting to the content or other comments. Despite this exposure to diverse sources, we argue that user behavior on Naver News likely mirrors media outlet-specific readership tendencies, driven by frequent exposure to content from preferred news media. For example, users can selectively subscribe to or bookmark preferred media, thus prioritizing these sources on their front page. However, this does not imply exclusive exposure to subscribed content. Additionally, users who do not subscribe to specific media may still exhibit selective attention towards articles aligning with their political leanings, influenced by factors such as media outlet names or headlines. Therefore, we assume that user reactions primarily represent the viewpoints of user groups whose political ideologies align with those of the news media. However, there is also likely a smaller, yet significant, group of users engaging in dissenting discourse. These users offer rebuttals or counterarguments that represent opposing viewpoints~\cite{buttliere2017reading}, which will be further explored in the following sections. 

\begin{figure*}[ht] 
	\includegraphics[width=\textwidth]{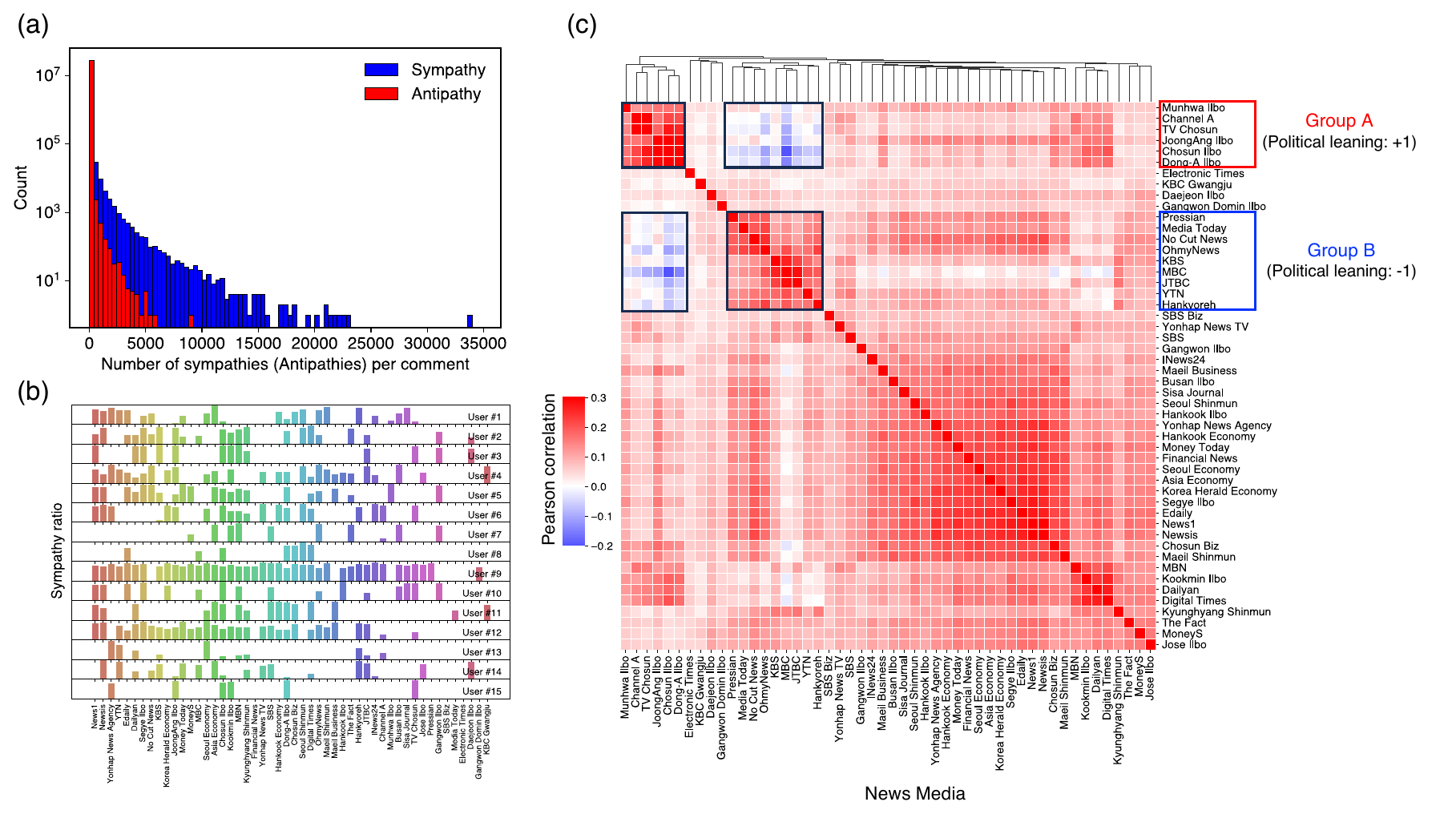}
	\caption{Characterization of political leaning of news presses based on comment-response relation between user groups. (a) The distribution of sympathy and antipathy response per comment. (b) Sympathy ratio distribution over 50 news presses. (c) Correlation matrix based on individuals' sympathy ratio distribution. This result demonstrates that correlation between news media based on sympathy ratio uncovers two conflicting media groups with opposing ideologies.}
	\label{FIG:response_statistics}
\end{figure*}

Based on the above assumption, we first examine whether we can find groups of news media with opposing opinions or ideologies from the observation of user comments and the responses received by other users. Especially, we utilize the proportion of sympathies and antipathies of comments over various news media. Figure~\ref{FIG:response_statistics}(a) illustrates the distribution of the number of sympathies and antipathies that comments of users received across all articles. The comments of news articles in the dataset received an average of 5.95 sympathies and 1.34 antipathies, indicating that a comment receives 4.44 times more sympathies than antipathies on average. 

If a user's comments in articles related to certain issues are likely to receive many sympathies from users in some news media but receive relatively more antipathies in the other news media, one might anticipate that the two news media groups and the corresponding readers hold opposing stances to the user regarding the topic. And if such different response patterns are observed in the comments from a large number of users during the election period, one can expect that these two news media groups have different political ideologies.

To characterize distinct responses to a user's comments from various media, we measure the ratio of sympathies and antipathies that a user received from different news media.
If a user $i$ receives $n_s^{p,m}$ number of sympathies and $n_a^{p,m}$ number of antipathies on a comment they have written to an article $p$ published from a news media $m$, the sympathy ratio of the comment is defined as $\rho_i^{p,m} \equiv n_s^{p,m}/(n_s^{p,m}+n_a^{p,m})$. Then, the sympathy ratio corresponding to news media $m$ of the user $i$ is calculated as the average sympathy ratio of each participated article, denoted as $\rho_i^{m}\equiv \left<\rho_i^{p,m}\right>_p$, where $\left< \cdots \right>_p$ signifies the average over the articles for which the user wrote comments.

To examine the macroscopic pattern of comment responses across different news media, we particularly focused on top 50 news media with largest article volumes. These 50 media produced 98.7\% of total articles within the dataset. To obtain statistically meaningful result, we also filtered out users with few number of comments. We extracted users who had written at least 10 comments to the articles in the 50 news media, where each of the comments has the sum of sympathies and antipathies more than or equal to 10. This resulted in a total of 2,344,682 comments from 67,455 users.

Figure~\ref{FIG:response_statistics}(b) displays the average sympathy ratio $\rho_i^{m}$s of 15 randomly sampled users across 50 news media. Considering a user's sympathy ratio over the news media as a vector of 50 elements, the sympathy ratios for entire users and news media can be represented as a matrix of dimensions 67,455 $\times$ 50. We then measured the pairwise correlation of sympathy ratios of users between news media resulting in a 50 $\times$ 50 correlation matrix, where each element depicts the correlation of responses of two news media. Figure~\ref{FIG:response_statistics}(c) exhibits the correlation matrix with a dendrogram showing the hierarchical-clustering structure of news media. 

Notably, while most news media' response behaviors exhibit positive correlation, there is a well-marked off-diagonal area with negative correlation values in Fig.~\ref{FIG:response_statistics}(c). We confirmed that this intergroup contrast remains consistent even when using non-parametric correlation measures such as Spearman and Kendall rank correlations, further supporting the robustness of the observed group division. The two groups of news media demonstrate 
relatively high positive intracluster correlations in response patterns compared to other news media and show negative intercluster correlations. The first group of news media (Group A) comprises six publishers: `Dong-A ilbo', `Chosun ilbo', `JoongAng ilbo', `Munhwa ilbo', `Channel A', and `TV Chosun'. The second group (Group B) consists of `Pressian', `Media today', `No Cut News', `OhmyNews', `Hankyoreh', `KBS', `MBC', `JTBC', and `YTN'. While there is no distinct and explicit concept of conservative or liberal news, the Group A is typically composed of news media considered conservative publishers, while the Group B consists of publishers representing liberal perspectives and broadcasting companies. 
From this perspective, utilizing our approach has the advantage of identifying two news media groups with contrasting patterns of user engagement, irrespective of somewhat ambiguous prior knowledge about conservative or liberal orientations of news media.
Based on our empirical observation of two groups of news media exhibiting contrasting comment response behaviors, we assigned a political leaning of +1 to the news media of Group A and -1 to the Group B. We note that we use the term ``political leaning'' here because our focus is on news articles specifically addressing the presidential election.

\subsection{Bias and polarization of comment activities of users in news portal}

In this section, we define the political leaning of individual users based on their comment activities across various news media and explore the ideological bias and polarization of the news portal users. Particularly, we focus on the two groups of news media with different political leaning, in terms of user response, as observed in the previous section. These two groups of media covers 21.8\% of the entire articles and 33.4\% of user comments of the dataset. 

Given the two groups of news media with contrasting political leaning (Group A: $+1$, Group B: $-1$) An individual user's political leaning is defined based on their comment activities. As we described in the previous section, we assume that users are more likely to leave comments on articles from news media that has similar political perspective~\cite{jurkowitz2020us, stroud2018american, frimer2017liberals, iyengar2008selective}. Defining individual leanings based on user participation to contents such as comments or sharing of online contents from media with predefined political leaning has been similarly conducted in much of the literature~\cite{cinelli2021echo, cota2019quantifying, hohmann2023quantifying}. 

Following the method suggested by Cota \textit{et. al.}~\cite{cota2019quantifying}, we define individual political leaning as follows. 
If a user $i$ writes a comment on $j$th article, and the corresponding news media's political leaning is $c_j\in[-1,1]$, the user's participation on news articles can be described as follows:
\begin{equation}
C_i = \{c_1, c_2, \cdots, c_{n_i}\}.
\end{equation}
Here, $n_i$ represents the total number of comments from user $i$, and $C_i$ denotes the list of political leanings of the news media to which the articles commented on by user $i$ belong.
Then the individual leaning $x_i$ is defined through the mean of $C_i$:
\begin{equation}
x_i \equiv \frac{\sum_{j=1}^{n_i} c_j}{n_i}.
\end{equation}

\begin{figure} 
	\includegraphics[width=\textwidth]{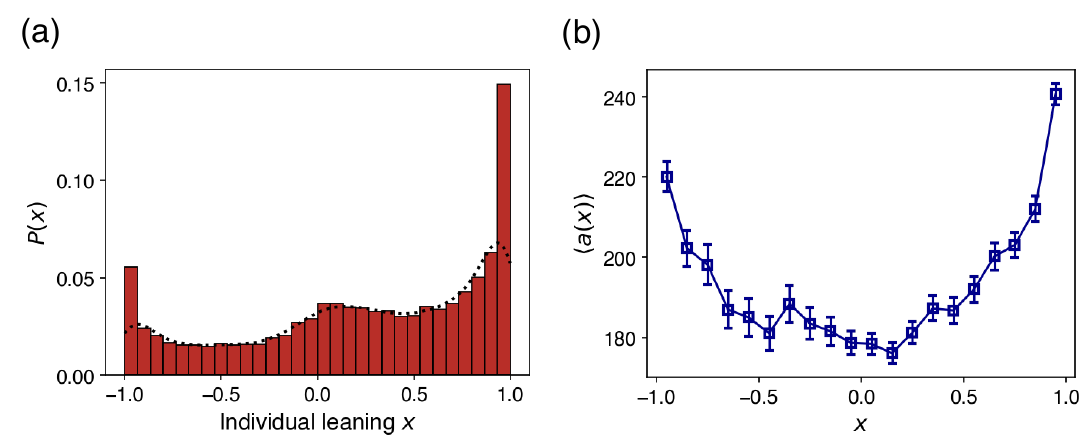}
	\caption{(a) Distribution of individual leanings $x$ among users. (b) Average comment activity $\left < a(x) \right >$ across different political leanings $x$. 
It is clearly shown that both (a) the political leaning of individual and (b) the average comment activity are peaked at the two extreme ends ($x=\pm 1$). }
	\label{FIG:individual_leaning}
\end{figure}

The distribution of number of comments made by a user is highly skewed, where most users produce very few comments (e.g., 75.2\% of users wrote less than 10 comments). Therefore, we focused on active users who wrote at least 100 comments during the election period period (189 days), resulting in 21,461 users who produced 46.4\% ($N=4,327,125$) of entire comments in the articles of two groups of news media.  

The distribution $P(x)$ of the individual leaning $x_i$ of the active users is displayed in Figure~\ref{FIG:individual_leaning}(a). Notably, the two clearly separated peaks in $P(x)$ suggest that users' political leanings fall into two extreme groups at both sides near $x=-1$ and $x=+1$. About 67.61\% of the users were found to have the conservative political leaning of $x > 0$ while 32.05\% of the users have the liberal political leaning of $ x < 0$. Despite Group B media producing more articles ($N=33,925$) compared to Group A ($N=23,270$), it was observed that users with $x > 0$ generated more comments ($N=2,754,859$) than those with $x < 0$ ($N=1,572,266$), suggesting a higher level of activity among users with conservative leanings ($x > 0$) on the Naver News platform.


Next, we measure the average user activity, denoted as $\langle a(x) \rangle$, where $a(x)$ is defined as the number of comments written by an individual user with political leaning $x$. For each value of $x$, we consider all users with political leaning $x$, record their respective numbers of comments $a^{(i)}(x)$, and compute the simple average across these users: $\langle a(x) \rangle = \frac{1}{N_x} \sum_{i=1}^{N_x} a^{(i)}(x),$
where $N_x$ is the number of users with political leaning $x$, and $a^{(i)}(x)$ denotes the number of comments written by user $i$. As shown in Fig.~\ref{FIG:individual_leaning}(b), the resulting $\langle a(x) \rangle$ exhibits a bimodal distribution, with peaks at the extreme values of $x=-1$ and $x=1$, similar to the pattern observed in Fig.~\ref{FIG:individual_leaning}(a).

In summary, our results reveal that the commenting behavior of users on the Naver News portal is highly biased toward both extreme ends, as indicated by the bimodal shape of $P(x)$ in Fig.~\ref{FIG:individual_leaning}(a). Furthermore, individuals with extreme political leanings are inclined to be more active participants on the platform as shown in Fig.~\ref{FIG:individual_leaning}(b) for $\langle a(x) \rangle$, suggesting that users with extreme views are even more prominent on the platform than those with moderate leanings.

\begin{figure*} 
	\includegraphics[width=\textwidth]{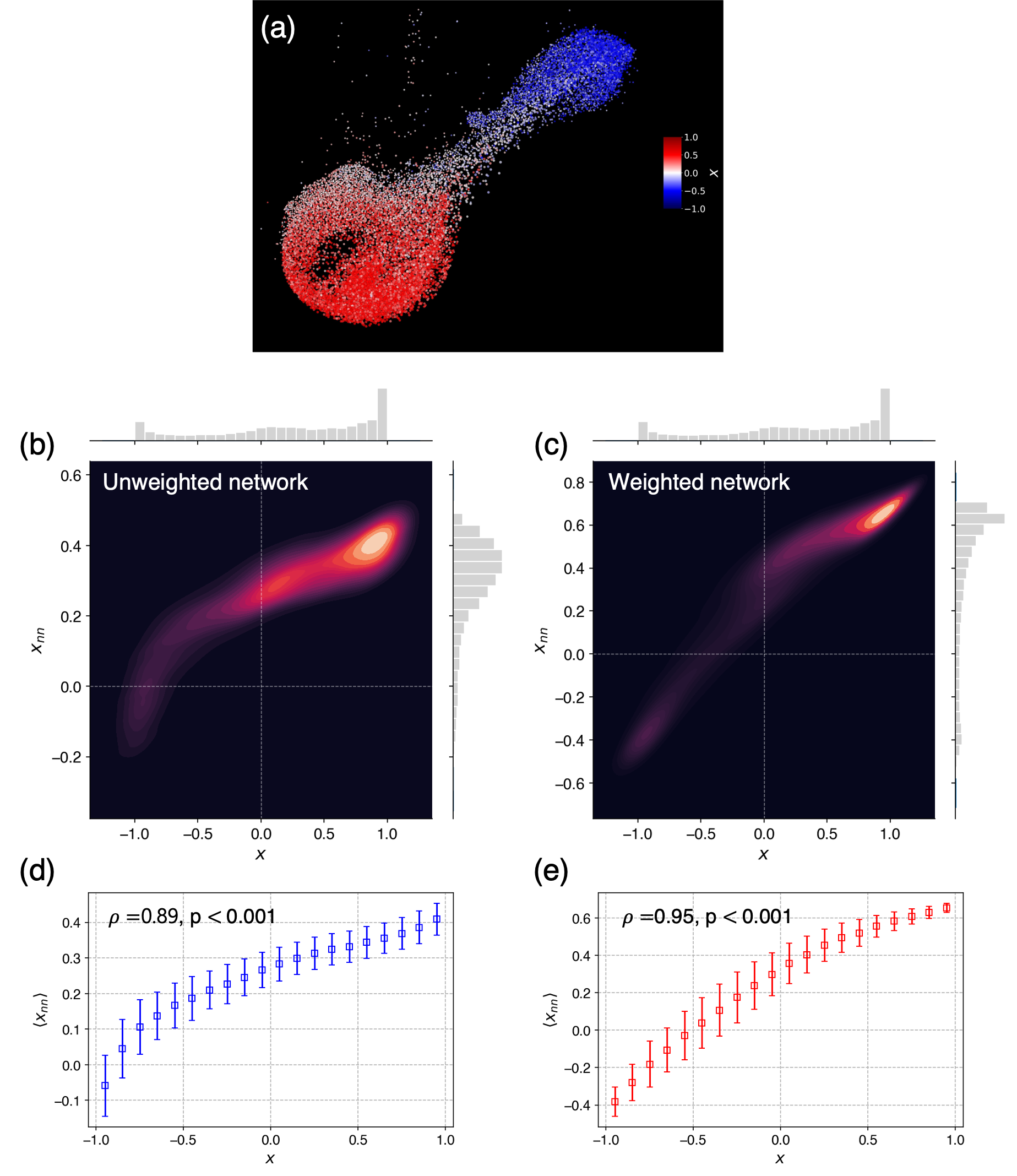}
	\caption{(a) Visualization of the co-commenting interaction network, comprising $N$=21,461 users, with users' individual political leanings $x$ color-coded from blue ($x=-1$) to red ($x=+1$). Node positions were determined using a force-layout algorithm, considering only the top 2\% of the total links (3,925,531 out of 355,467,509) based on link weight for visualization purposes. In the visualization, only nodes are shown. (b) and (c) Joint distributions of individual political leaning $x$ among users and the average leaning of their nearest neighbors $\left < x_{nn} \right >$ for (b) the unweighted and (c) the weighted co-commenting interaction networks. (d) and (e) show the average $\left < x_{nn} \right >$ across political leanings $x$ that corresponds to (b) and (c), respectively. The weighted version of the interaction network demonstrates a clearer separation of two clusters in (c) and higher assortativity in political leaning in (e).}
	\label{FIG:x_vs_xnn}
\end{figure*}

\subsection{Echo chamber of like-minded users in the co-commenting interaction network}

The observation that the political leanings of individuals are polarized into two extreme values prompts us to pose another inquiry: Do users with similar political leanings interact more actively with like-minded individuals? Since the polarization of individual leanings does not necessarily imply gathering of similar users in their communication structure, we investigate the interaction network between users based on co-commenting on the same news articles. 

The interaction network of users is constructed by considering comments. Here, we assume that users who commented on the same article share an interest (either the intention is positive or negative) in that specific article and are exposed to each other's opinions through the contents of their comments. In this network, two users are connected if they co-commented on at least one news article. We generate both the unweighted version of the interaction network, where all links have equal weights, and the weighted interaction networks, where links between two users are weighted by the number of articles on which they both commented. Figure~\ref{FIG:x_vs_xnn}(a) shows the visualization of the weighted version of interaction network, comprising 21,461 users who wrote at least 100 comments, with their individual political leanings are color-coded from blue ($x = -1$) to red ($x = +1$). 

To determine if there is correlation between the political leaning of a node (individual) and its neighbors, we plot the average leaning of neighbor nodes $\left< x_{nn} \right >$ across different levels of individual leaning $x$. Figures~\ref{FIG:x_vs_xnn}(b) and (c) show that there is a strong correlation between values of $x$ and $\left< x_{nn} \right >$ in both of the unweighted and the weighted networks. The Pearson correlation between $x$ and $\left< x_{nn} \right >$ is $\pi=0.89$ ($p < 0.001$) for the unweighted network and $\pi=0.95$ ($p < 0.001$) for the weighted network [Figs.~\ref{FIG:x_vs_xnn}(d) and (e)], indicating that the nodes consisting the interaction network are highly assortative in terms of their political leaning.

Compared to the unweighted network, the weighted interaction network exhibits a clearer separation of users, as demonstrated by two dense clusters in the joint probability distribution of $x$ and $\left< x_{nn} \right >$ in Fig.~\ref{FIG:x_vs_xnn}(c). This result indicates that users on Naver News portal not only exhibit a biased distribution of political perspectives but are also more likely to interact with and be exposed to like-minded users, which is a significant sign of an echo chamber effect.

\subsection{Analyses of comment reaction unveil asymmetric affective polarization between two user groups}

In the previous subsection, we have demonstrated the existence of two clusters of users with a similar political leaning in the co-commenting interaction network. While this indicates a structural polarization between two user groups, it does not necessarily imply the presence of emotional conflict between the two groups. Therefore, we investigate the aggregated affection toward each ideological group by analyzing user reactions to comments written by users with a specific individual leaning, from news media of distinct political leanings.  Particularly, we measure the `replies', `sympathies', and `antipathies' that a certain comment of a user received from other users in two media groups A and B with diverging political leanings. In a news article page, when a user makes a comment on a news article, other users can reply or vote a `sympathy' or `antipathy' to the comment. Although direct information about the respondents such as who replied to a comment or who voted for `sympathy' or `antipathy' is not included in our dataset, the aggregate response statistics (the number of replies, sympathies, and antipathies) are available. However, it is important to acknowledge that these aggregated response statistics serve as proxies for ideological and emotional stance, and may not fully capture the complex emotional states or underlying motivations behind individual user interactions.

\begin{figure*} 
	\includegraphics[width=\textwidth]{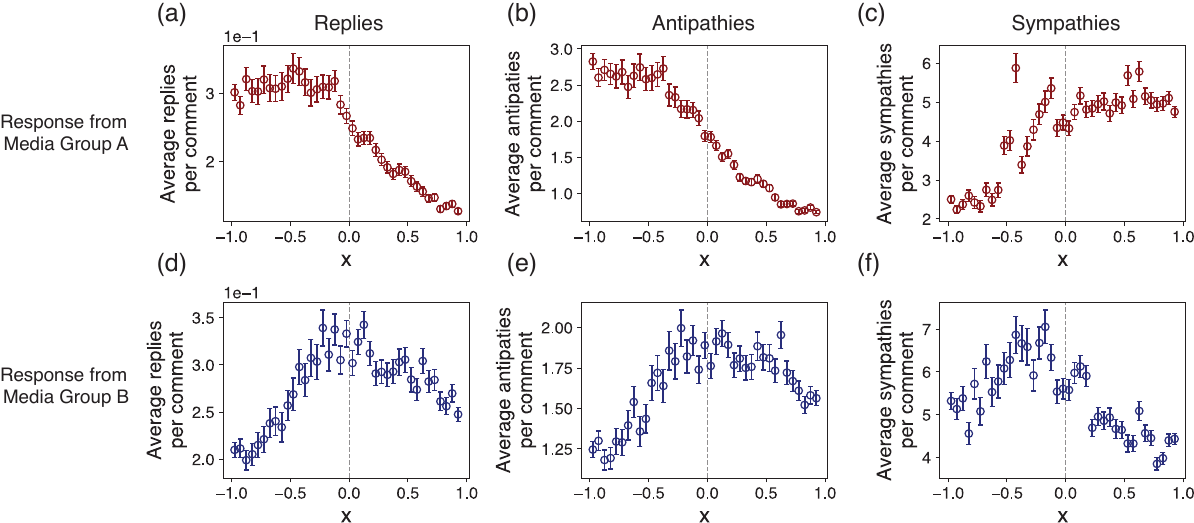}
	\caption{The relationship between the individual leaning $x$ of commenters and the corresponding responses from media groups A and B. We measure the average responses in three different ways: [(a), (d)] replies, [(b), (e)] antipathies, and [(c), (f)] sympathies for two media groups: [(a), (b), (c)] Group A and [(d), (e), (f)] Group B. While the response pattern from Group A shows a linear dependence on $x$, user responses in Group B exhibit a more mixed pattern, indicating an asymmetric response pattern between the two groups.
}  
	\label{FIG:x_vs_reactions}
\end{figure*}

Therefore, instead of focusing on individual interactions, we examine the macroscopic behavior of comment responses. From the two media groups (Group A and Group B) with largely opposing political leanings, we measure how the reactions to comments generated by users in each group varies with the individual political leaning of users who originally wrote the comment.
Figure~\ref{FIG:x_vs_reactions} represents how users in two media groups reacted to the comments of users with different levels of individual leaning $x$ based on average number of replies, sympathies, and antipathies per comment. Interestingly, the shape of three responses from two groups exhibit distinctive patterns.

First, in case of Group A (media with conservative political leaning $x=+1$), all three user responses are strongly correlated with individual leaning $x$ of users who wrote the comments as displayed in Figs.~\ref{FIG:x_vs_reactions}(a)-(c). As can be clearly seen in Figs.~\ref{FIG:x_vs_reactions}(a) and (b), both average number of replies and antipathies per comment show a decreasing tendency as $x$ is increased. On the contrary, average sympathies per comment tends to increase with $x$ [see Fig.~\ref{FIG:x_vs_reactions}(c)]. This result supports that users (respondents) in Group A have a clear preference for in-group members and hostility towards others, which is a signature of affective polarization. 

While the response patterns from Group A show an almost linear dependence on individual leaning $x$, user responses in Group B exhibit a different pattern. Once again, the average number of replies and antipathies are highly correlated. However, in contrast to Group A, both average replies and antipathies increase for $x < 0$ and then decrease for $x > 0$, exhibiting inverted U-shape graphs. Therefore, the users in Group B
are likely to respond more actively and negatively to the users with intermediate political leanings. Rather, they appear to have chosen a strategy of ignoring and not reacting to users with opposing opinions ($x$ close to $+1$).

However, in terms of expressing preferences, the average sympathies from users in Group B are found to increase as the commenter's individual leaning ($x$) decrease from 1 to -1, peaking before $x\approx-0.5$. This suggests that Group B users generally favor comments from users with similar political orientations ($x<0$). However, their sympathy diminishes for comments made by individuals with extreme views near $x = -1$, indicating a reduced preference for possibly  more radical views, even within their own ideological group. Overall, both media groups show a preference for in-group members, while they exhibit distinct expressions of hostility towards out-group members, indicating an asymmetric affective polarization found in comment responses.

\begin{figure*} 
	\includegraphics[width=\textwidth]{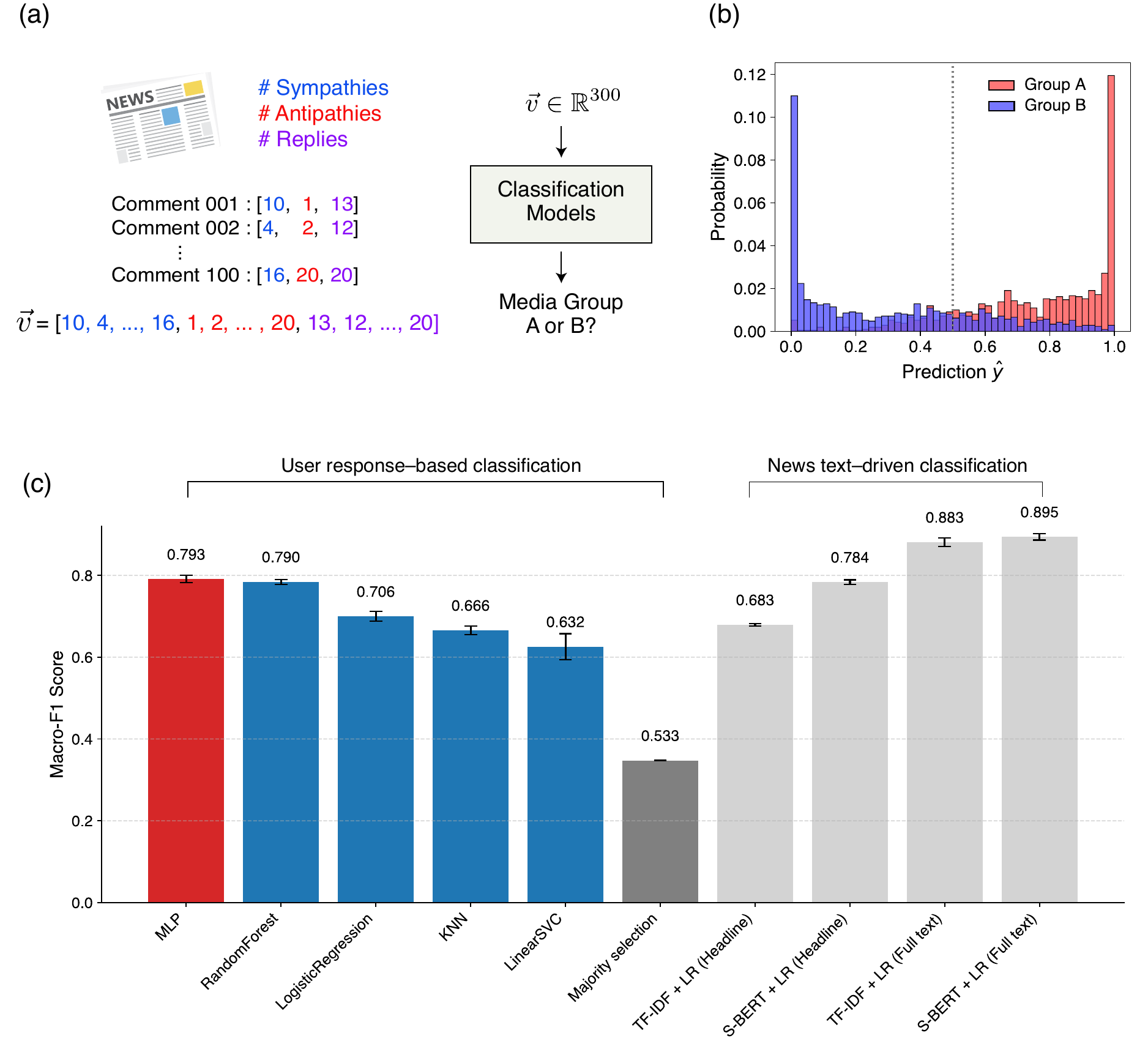}
	\caption{Classification task aimed at categorizing news articles into different media groups based on response statistics (number of sympathies, antipathies, and replies). 
 (a) Illustration of the media classification process: For each news article, the top 100 comments with the most responses were collected. The numbers of sympathies, antipathies, and replies for each comment were concatenated into a 300-dimensional vector. These vectors were used to train classification models to predict the political leaning of the news media outlet associated with each article.
 (b) Prediction results of the classification model using a multi-layer perceptron (MLP) on test set articles. The model was trained to output 1 for Media Group A and 0 for Media Group B. 
(c) Performance comparison across classification models using the macro F1-score. The first six models represent classification based on user response patterns, while the last four models are based on textual features (news headlines and full text) for comparison. The MLP achieved the highest F1-score (0.793), followed by the Random Forest model (0.790). These results highlight that distinct response patterns on comments from collective users provide meaningful signals for distinguishing media groups with different ideologies, outperforming models that rely on news headlines.}
	\label{FIG:classification}
\end{figure*}

\subsection{Classifying articles into different media groups based on collective user response statistics}

Our results highlight an asymmetric affective response pattern between user groups with diverging political leanings. This suggests the possibility that different political groups may adopt distinct communication strategies. To support this hypothesis, we test whether we can predict the political leaning (+1 or -1) of an article using only simple comment response statistics (number of sympathies, antipathies, and replies) obtained from multiple comments.

Figure~\ref{FIG:classification}(a) illustrates the overall process of media classification task. We utilized 10,505 articles with more than 200 comments and divided the dataset into three sets: a training set (6,303), a validation set (2,101), and a test set (2,101). This process was repeated five times for 5-fold validation.  

For each article in the training set, we selected the top 100 comments with the most responses (the total sum of sympathies, antipathies, and replies). Subsequently, the numbers of sympathies, antipathies, and replies of these 100 comments were concatenated into a 300-dimensional vector. These vectors were then used to train various classification models to predict the political leaning of the news media to which an article belongs. We tested several models, including a Multi-layer Perceptron (MLP), Random Forest, Logistic Regression, K-Nearest Neighbor (K-NN), and Support Vector Machine (SVM), and compared their performance with a majority rule prediction baseline. For the MLP, we employed a neural network consisting of two hidden layers with a sigmoid output node, implemented using the \texttt{pytorch} library~\cite{paszke2019pytorch}. Other classification models were implemented using the \texttt{Scikit-learn} library~\cite{scikit-learn}, with hyperparameters selected based on performance on the validation set. Model performance was evaluated using the macro F1-score to account for potential class imbalance across two media groups.

We additionally conducted news text-driven classification using the same dataset to provide a comparative benchmark for the comment response-based classification models. We implemented two text-based classification models: (i) a TF-IDF vectorizer followed by a Logistic Regression classifier (TF-IDF + LR), and (ii) a Sentence-BERT (S-BERT) model combined with Logistic Regression (S-BERT + LR). For the S-BERT model, we employed the pretrained Korean language model \texttt{snunlp/KR-SBERT-V40K-klueNLI-augSTS}~\cite{kr-sbert}. Each model was trained using either the news headline alone or the full article text (headline concatenated with body text) as input. In the case of full text, we used up to 512 tokens to match the input size constraints of the S-BERT model. These text-based models serve as a baseline to assess the relative effectiveness of classification based solely on user comment response statistics.

Figure~\ref{FIG:classification}(b) demonstrates that the MLP successfully classifies articles from different media groups into extreme values of political leanings. Figure~\ref{FIG:classification}(c) exhibit the performance across different classification models. While the baseline F1-score using majority selection is only 0.533, the MLP achieves the highest accuracy macro F1-score at 0.793, followed by the Random Forest model at 0.790. All response-based classification models present the performance surpassing the baseline by a significant margin. 

In Fig.~\ref{FIG:classification}(c) we also compare the response-based classification models with text-driven models. Notably, the MLP trained solely on response statistics outperforms the headline-based text model (S-BERT + LR), which achieves an F1-score of 0.784. Although the models using the full article text achieve higher F1-scores (0.895 for S-BERT + LR, 0.883 for TF-IDF + LR), the fact that response-based models can match or exceed the performance of headline-based models is particularly noteworthy. This finding suggests that collective user behaviors, as expressed through comment-level interactions, capture distinctions between two media groups beyond textual features of the articles. Taken together with the results in Fig.~\ref{FIG:x_vs_reactions}, these findings support the notion that media audiences aligned with opposing ideologies engage in distinct communicative strategies, which are systematically reflected in their aggregated affective response patterns.

\section{Discussion and Conclusion}
This study has provided a comprehensive network analysis of news media polarization and user engagement patterns during the 2022 presidential election in South Korea, utilizing a large-scale dataset from the Naver News portal. Our investigation has unveiled several significant findings, including the clustering of news media based on user comment responses, the presence of bias and polarization in user comment activities, the existence of echo chambers within co-commenting networks, and the observation of asymmetric affective polarization between user groups.

Our key findings include the followings. First, our study suggests a novel approach to identify media groups with contrasting political leanings based on the user engagement pattern. We emphasize that we do not take actual contents of articles and comments into account at all, but use only numbers of replies, sympathies, and antipathies in our analyses. We found two distinct clusters of news media, differentiated by the sympathy and antipathy responses of users, indicating opposing political ideologies. This approach offers a novel perspective on understanding media landscape through user interactions, moving beyond traditional survey-based analysis on article contents. User analysis revealed a clear bimodal distribution of political leanings among active users, indicating a significant ideological bias and polarization among users in the news platform. Interestingly, users with extreme political views were found to be more active, highlighting the dominance of their opinions on the platform.

Additionally, our results revealed the presence of echo chambers within co-commenting networks, where users with similar political leanings predominantly interact within their own groups. This phenomenon was observed in both unweighted and weighted versions of interaction networks, indicating a strong assortativity based on political leanings. Furthermore, the examination of comment responses highlighted asymmetric affective polarization between user groups with diverging political ideologies. While a preference for in-group members was evident in both groups, their expressions of hostility and responsiveness toward out-group members differed depending on the media groups. Building upon the discovery of diverging response patterns in different media groups, we conducted a media classification task. The high predictability of classification models on media leaning of articles solely through basic comment response statistics highlights that aggregated user responses provide meaningful information about the group behavior to which the users belong. This finding supports the idea that different political groups may form and employ distinct communication tactics.

However, this study is subject to several limitations that future research could aim to address. First, the analysis was limited to data from a single news portal during a specific time frame, namely the presidential election period. An expansive investigation across various online news platforms and over more extended periods would likely yield a more comprehensive understanding of user dynamics in general online news contexts. 
While our study focuses on aggregated affective responses during the overall election period, it does not explicitly consider the temporal heterogeneity of user sentiment across different stages of the election cycle. Prior research has shown that user attention and emotional responses vary dynamically depending on the timing and nature of events~\cite{gao2015uncovering}. For example, attention intensity and sentiment expression may peak around key moments such as debates or voting day. Incorporating time-series clustering techniques could enable future work to identify phase-specific patterns in affective polarization~\cite{li2017reasoning}, thereby offering a more nuanced understanding of user behavior throughout the electoral process.

Second, the study's reliance on observed behaviors such as comment sympathies and antipathies may lead to potential misinterpretations of user intentions, as it does not directly assess the underlying motives or sentiments behind these actions. Future research could build on our approach by incorporating direct analyses of the content of user comments, enabling a richer understanding of affective expressions and ideological stances beyond reaction patterns alone. Recent work has demonstrated the value of combining content-based and interaction-based signals to capture nuanced forms of stance endorsement and opposition within polarized communities~\cite{ebeling2022analysis}.

Third, while our binary classification of media groups successfully captures the dominant structure of polarization observed in user responses, it inevitably simplifies the more complex and nuanced spectrum of political ideologies among news outlets. A portion of media outlets and users exhibited response patterns that did not align neatly with either cluster, reflecting the presence of more moderate or mixed-ideology stances. This suggests that political diversity in the media landscape extends beyond a strict dichotomy. Although our focus on the two polarized groups allows for a clear illustration of affective and structural polarization, future research could explore finer-grained classifications or continuous models of political ideology to better capture the full heterogeneity of media and user behaviors.

Fourth, an important aspect that deserves further discussion is the potential role of the Naver News algorithm in shaping user behavior and reinforcing the clustering of like-minded users. On the platform, the visibility of news articles is not user-independent; rather, it is mediated by an internal recommendation algorithm that curates each user's feed based on prior engagement history and subscriptions. Although we do not have access to the exact mechanisms of this algorithm, it is widely recognized that such algorithmic filtering can influence which articles users are more likely to encounter~\cite{gonzalez2023asymmetric, guess2023social}. This personalization may amplify selective exposure and homophily, thereby contributing to the formation of echo chambers and the polarization patterns observed in our study~\cite{pariser2011filter}. While our analysis focuses on user comment behavior without direct access to article exposure data, we acknowledge that algorithmic curation may partially shape the behavioral patterns we report. Future work leveraging more detailed information on content exposure mechanisms could help disentangle the respective roles of algorithmic design and user preference in shaping political polarization on online news platforms.

For future research, incorporating a wider array of data sources and direct user interaction information, and examining the impact of algorithmic biases could enrich the understanding of media polarization. Investigating the complexities of the political spectrum of news media and the psychological drivers behind user engagement would also provide deeper insights into the societal impacts of media polarization. 

Although our study focuses on a South Korean news platform, the proposed media clustering methodology and the framework used to assess collective user response patterns are broadly generalizable to other contexts. Because our media clustering method relies solely on user engagement patterns rather than predefined political categorizations, it can be applied to diverse online platforms where users interact with content, such as \textit{Reddit}, \textit{Quora}, and \textit{YouTube}. This flexibility suggests that our approach may serve as a versatile tool for uncovering group structures and affective polarization across a wide range of online communities beyond traditional news ecosystems. Thus, our framework offers a promising foundation for comparative studies of polarization and collective behavior across different sociocultural and platform contexts.

\section*{Acknowledgement}
This work was supported by the Ministry of Education of the Republic of Korea and the National Research Foundation of Korea (NRF-2021S1A5A2A03068831). H.J. acknowledges support from the National Research Foundation of Korea (NRF RS-2025-00514776).

\bibliographystyle{elsarticle-num}

\bibliography{main}

\end{document}